\documentclass[letterpaper,twocolumn,prl]{revtex4-1}
\usepackage[utf8]{inputenc}
\usepackage{fourier}            
\usepackage{times}              
\usepackage{inconsolata}        
\usepackage{latexsym,amsmath,amssymb,bbm,amsthm}
\newcommand\ket[1]{|#1\rangle}
\newcommand\bra[1]{\langle #1|}

\renewcommand{\openone}{\ensuremath{\mathbbm 1}}
\newcommand\ri[2]{S\left({#1} \,|\kern-0.05em|\, {#2}\right)}
\usepackage{hyperref}

\date{\today}

\begin{document}
\author{M. Boyer}
\affiliation{Département IRO, Université de Montréal, Montréal H3C 3J7, Canada}
\title{Zero discord implies classicality}
\begin{abstract}
The ``classical-quantum'' (\emph{cq}) discord of a bipartite state $\rho^{AB}$ is the smallest difference
between the mutual information $S(\rho^{A:B})$ of $\rho$ and that of $\rho$ after a measurement
channel is applied on the $A$ system. Relating zero discord to the strong subadditivity of the Von Neumann
entropy, Datta  
proved that a state has zero \emph{cq} discord iff and only if it can be written
in the form $\sum_i p_i\ket{i}\bra{i}\otimes \rho^B_i$ for $p_i$ a probability distribution,  $\ket{i}$ a basis of the $A$ system and  $\rho^B_i$ states of the $B$ system.
We provide a simple  proof of that same result  using directly a theorem of Petz on channels that leave 
unchanged the relative entropy of two given states. 
\end{abstract}

\maketitle
Various measures have been proposed to quantify
the classicality or equivalently the quantumness in
a multipartite state.
One of them, ``discord'', is the theme of this note.
Limiting ourselves to a bipartite system $AB$, given an orthonormal basis $\mathsf{A} = \big\{\ket{a}\big\}_{a\in A}$ of
$\mathcal{H}_A$, 
any state $\rho^{AB}$, i.e. density operator, can be represented by a block
matrix with blocks  $\rho^B_{aa'}$, corresponding to $\rho^{AB} = \sum_{aa'}\ket{a}\bra{a'}\otimes \rho^B_{aa'}$.
The state is said to be ``classical-quantum'' (cq) if there 
is such a basis for which the matrix is block diagonal
i.e $\rho^B_{aa'} = 0$ if $a\neq a'$. 
The operator $\mathcal{D}_{\mathsf{A}}= \sum_{a\in {A}} (\ket{a}\bra{a}\otimes\mathbf{1}_B)\rho(\ket{a}\bra{a}\otimes\mathbf{1}_B)$ which replaces  off diagonal blocks 
  by a $0$ block is
a quantum channel that acts independently on the $A$ and the $B$ system (it is the identity on $B$). 
It is known that for such channels $\mathcal{E}$ holds the inequality 
$I(\mathcal{E}(\rho)^{A:B}) \leq I(\rho^{A:B})$ where
$I(\rho^{A:B})$, the \emph{mutual information}, is $S(\rho^A) + S(\rho^B)-S(\rho^{AB})$ and 
$S(\rho) = - \mathrm{tr}\big[\rho \lg(\rho)\big]$. It is the infimum of $I(\rho^{A:B}) - I(\mathcal{D}_{\mathsf{A}}(\rho)^{A:B})$ over all bases $\mathsf{A}$ that
was
called ``discord'' by Ollivier and Zurek \cite{OZ} and that we shall call \emph{qc} ``classical-quantum'' discord
(provided the dimension of $\mathcal{H}_A$ is properly chosen).
It is clear that if $\rho$ is classical-quantum, 
then the \emph{qc} discord is $0$.
The converse is not obvious. The argument in \cite{OZ} appears to lead nowhere (cf appendix).
Datta \cite{Datta2010} gave a proof  that zero discord implies classical-quantum 
using a result of Hayden et al. \cite{HJPW} on
the structure of states which satisfy strong subadditivity of quantum entropy with equality.
His definition of discord however looks more general than the above since he 
optimizes over all channels defined by rank $1$ POVM's on the $A$ system instead of
complete projective measurements. Nevertheless a rank $1$ POVM with outputs in $M$ is
equivalent to a unitary embedding of $\mathcal{H}_A$ into $\mathcal{H}_M$ with
basis the $\ket{m}$  and we can simply apply a projective measurements in $\mathcal{H}_M$.
In fact, we can always choose $M$ of size at most $d_A^2$ where $d_A = \dim\mathrm{supp}(\rho^A)$
so that we need only choose $\mathcal{H}_A$ of dimension $d_A^2$ and work with
block matrices.
Details are to be found in the appendix where it is also shown that there is always
a basis corresponding to the  discord. The following theorem thus implies
that if the \emph{cq} discord is $0$, the state is classical-quantum. The approach is similar to that of Piani
et al \cite{HJPW} for ``classical-classical'' states.

\medskip

\noindent\textit{\textbf{Theorem}}\quad  
\textit{Let $\mathcal{D}=\mathcal{D}_\mathsf{A}$. 
If\/ $I(\mathcal{D}(\rho)^{A:B}) = I(\rho^{A:B})$ then 
$\rho$ can be block diagonalized in some basis of $\mathcal{H}_A$.
}
\begin{proof}
The equality $I(\mathcal{D}(\rho)^{A:B}) = I(\rho^{A:B})$ is equivalent to
$\ri{\mathcal{D}(\rho^{AB})}{\mathcal{D}(\rho^A\otimes\rho^B)}=\ri{\rho^{AB}}{\rho^A\otimes\rho^B}$.
A theorem of Petz states that if a channel $\mathcal{E}$ is such that
$\ri{\mathcal{E}(\rho)}{\mathcal{E}(\sigma)} = \ri{\rho}{\sigma}$ then there exists $\widehat{\mathcal{E}}$ such
that $\widehat{\mathcal{E}}\mathcal{E}(\rho)=\rho$ and moreover $\widehat{\mathcal{E}}(Y) = \sigma^{1/2}\mathcal{E}^*\left(\big(\mathcal{E}(\sigma)\big)^{-1/2}Y\big(\mathcal{E}(\sigma)\big)^{-1/2}\right)\sigma^{1/2}$
where $\mathcal{E}^*$ is the adjoint of $\mathcal{E}$.
Letting $\rho^{AB} =\sum_{aa'}\ket{a}\bra{a'}\otimes\rho^B_{aa'}$,
$p_a = \mathrm{tr}\big[\rho^B_{aa}\big]$, $p_a\rho^B_a = \rho^B_{aa}$ and 
$\sigma=\rho^A\otimes\rho^B$ gives $\mathcal{D}(\sigma) = \sum_a p_a \ket{a}\bra{a}\otimes \rho^B$
and $\mathcal{D}(\rho^{AB}) = \sum_a p_a\ket{a}\bra{a}\otimes \rho^B_a$.
It follows that $(\mathcal{D}\sigma)^{-1/2} = \sum_{a\in A} p_a^{-1/2}\ket{a}\bra{a}\otimes \rho_B^{-1/2}$ and
\begin{equation}\label{frompetz}
\rho^{AB}=\widehat{\mathcal{D}}(\mathcal{D}(\rho^{AB})) = \sum_{a\in A}\rho_A^{1/2}\ket{a}\bra{a}\rho_A^{1/2}\otimes \rho^B_a
\end{equation}
and $(\ket{a}\bra{a}\otimes\mathbf{1}_B)\rho^{AB}(\ket{a}\bra{a}\otimes\mathbf{1}_B) 
= \sum_{a'\in A} \big|\bra{a}\rho_A^{1/2}\ket{a'}\big|^2 \rho^B_{a'}= \rho^B_{aa}=
p_{a}\rho^B_{a}  
$;
if $p_a\neq 0$ then
\begin{align}
\rho^B_{a} &= \sum_{a'\neq a}p_{a'}^{\phantom{B}} \rho^B_{a'} && p_{a'} = \frac{\big|\bra{a}\rho_A^{1/2}\ket{a'}\big|^2}{p_a-\big|\bra{a}\rho_A^{1/2}\ket{a}\big|^2}\label{preconvcomb}
\end{align}
 so that 
 each diagonal block is a convex combination of the others.
Thus, for all the extremal states $\rho^B_{a}$ of the convex hull of the $\rho^B_a$, 
$\big|\bra{a'}\rho_A^{1/2}\ket{a}\big|^2 = 0$ if $\rho_{a'} \neq \rho_{a}$. If we consider the non extremal states, the  extremal ones do not appear
in their convex combination 
 \eqref{preconvcomb} and we may apply the same argument to their convex hull, and so on, eventually getting that $\bra{a}\rho^{1/2}\ket{a'} = 0$ if  
 $\rho^B_a \neq \rho^B_{a'}$. 
 Grouping together the $a$ with equal $\rho^B_a$ gives
 a partition $A_1,\ldots, A_k$ of $A$ and Eq. \eqref{frompetz} becomes
 \begin{align*}
 \rho^{AB} = \sum_{i=1}^k P_i \otimes \rho_{a_i} && P_i = \rho_A^{1/2}\left(\sum_{a\in A_i}\ket{a}\bra{a}\right)\rho_A^{1/2}
 \end{align*}
with $\bra{a'}P_i\ket{a}=0$ if either $a' \in A \backslash A_i$ or $a\in A\backslash A_i$. Letting
   $\mathcal{H}^A_i = \mathrm{Span}\big\{\ket{a}: a\in A_i\big\}$,
  that implies that $\mathrm{Supp}(P_i)\subseteq \mathcal{H}^A_i$; the supports of the $P_i$ being
  pairwise orthogonal, the $P_i$ can be simultaneously diagonalized, letting $\rho^{AB}$  block diagonal.
\end{proof}

\noindent We thank Kavan Modi for useful comments.
 
%
 
\appendix
\section{Optimizing over all measurement maps}
In this section, if $\mathcal{E}$ is defined on the first system only then, when applied also on $B$, it is $\mathcal{E}\otimes\openone_B$ that
is meant.

A POVM $\mathbf{M}=\{M_m\}_{m\in M}$ is a family of operators 
such that $0\leq M_m \leq \mathbf{1}_A$ and
$\sum_{m\in M} M_m = \mathbf{1}_A$. For any measurement procedure on $\mathcal{H}_A$ with outputs in $M$, there is a POVM
such that the probability of output $m$ given the state $\rho$ is $\mathrm{tr}(M_m\rho)$ \cite{NC2000}.
Following \cite{PHH}, we call \emph{measurement map} associated to $\mathbf{M}$ the channel defined by $\mathcal{M}_{\mathbf{M}}(X) = \sum_{m\in M}\mathrm{tr}\big(M_m X\big)\ket{m}\bra{m}$. If $\rho$ is a state of $\mathcal{H}_A$ then
$\mathcal{M}_{\mathbf{M}}(\rho)$ is state of $\mathcal{H}_M$ with ``standard basis'' $\big(\ket{m}\big)_{m\in M}$.

A POVM $\mathbf{M}' = \{M'_{m'}\}_{m'\in M'}$ with outputs in $M'$ is a \emph{refinement} of $\mathbf{M} = \{M_{m}\}_{m\in M}$,
is a map $p:M'\to M$ s.t. $M_m = \sum_{m'\in p^{-1}(m)} M'_{m'}$.
$\mathbf{M}$ corresponds to returning the output $p(x')$ each time $\mathbf{M}'$ output $x'$ , i.e. $\mathbf{M}$ groups the outputs of $\mathbf{M}'$ using $p$, and sums their probabilities.
If  $\mathbf{M}'\succeq\mathbf{M}$ then there is a channel $\mathcal{E}$ s.t. $\mathcal{E}\circ\mathcal{M}_{\mathbf{M}'} = \mathcal{M}_{\mathbf{M}}$: Let $\mathcal{E}(X) = \sum_{m\in M} A_mX A_m^\dagger$ where $A_m = \sum_{m'\in p^{-1}(m)} \ket{m}\bra{m'}$;
$\sum_m A_m^\dagger A_m = \mathbf{1}_{\mathcal{H}_{M'}}$ and
$
\mathcal{E}\mathcal{M}_{\mathbf{M}'}(X) = \sum_{m\in M}\sum_{m'\in p^{-1}(m)} \ket{m}\mathrm{tr}\big(M'_{m'}X\big)\bra{m} =
\mathcal{M}_{\mathbf{M}}(X)
$

The set of POVMs with output set $M$ is convex. Since $\mathcal{M}_{\mathbf{M}}$ is linear in $\mathbf{M}$ and
the relative information is convex in its inputs,
any POVM with outputs in $M$ and for which $I(\mathcal{M}_{\mathbf{M}}(\rho)^{M:B})$ is optimal must be extremal \cite{ModiBrodutch}.
Letting $M_m = \sum_{n=1}^{d_m}\ket{e^m_n}\bra{e^m_n}$ be a spectral decomposition of $M_m$, $\mathbf{M}$ is extremal iff  the
$\sum_{m\in M} d_m^2$ operators $\ket{e^m_n}\bra{e^m_{n'}}$ for $1\leq n, n' \leq d_m$ are linearly independent \cite{Dariano}, so that
there is at most $d^2$ operators  $\ket{e^m_n}\bra{e^m_n}$ where $d=\dim\mathcal{H}_A$. Let $\mathbf{M}'$ be the refinement of $\mathbf{M}$ 
with those POVM elements;
$\mathbf{M}'$ is extremal, of rank one, with at most $d^2$ elements.
Moreover, since $\mathcal{M}_\mathbf{M}=\mathcal{E}\mathcal{M}_{\mathbf{M}'}$ for some  $\mathcal{E}$ and since
$I(\mathcal{E}(\mathcal{M}_{\mathbf{M}'}(\rho))^{M:B}) \leq I(\mathcal{M}_{\mathbf{M}'}(\rho)^{M':B})$, it follows
that the optimum mutual information is obtained considering only  rank $1$  POVMs indexed by a set $M'$ of size $d^2$.

It is however needed to prove that there is actually a POVM for which the optimum is realized. A POVM $\ket{e_m}\bra{e_m}$ of
rank $1$ on $\mathcal{H}_A$ defines an embedding $\imath:\mathcal{H}_A \to \mathcal{H}_M$ by
$\imath = \sum_{m\in M} \ket{m}\bra{e_m}$. An easy calculation shows that $\imath^\dagger\imath = \mathbf{1}_A$ if and only
if $\ket{e_m}\bra{e_m}$ is a POVM. Conversely, any embedding $\imath:\mathcal{H}_A\to\mathcal{H}_M$ is defined by
a POVM: from $\imath\ket{\phi} = \sum_m \ket{m}\bra{m}\imath\ket{\phi}$ it follows that $\bra{e_m} = \bra{m}\imath$.

Finally, the probability of measuring $m$ with the POVM defined by the $\ket{e_m}\bra{e_m}$ given $\rho$ 
is the same as the probability
of measuring $\ket{m}$ in the standard basis of $\mathcal{H}_M$ given the state $\imath\rho\imath^\dagger$.
The probability of measuring $m$ given $\rho$ is $\mathrm{tr}\big[\ket{e_m}\bra{e_m}\rho\big] = \bra{e_m}\rho\ket{e_m}$.
On the other hand, $\imath\rho\imath^\dagger = \sum_{mm'}\ket{m}\bra{e_m}\rho\ket{e_{m'}}\bra{m'} =
\sum_{mm'}\ket{m}\bra{m'}\bra{e_m}\rho\ket{e_{m'}}$ and the probability of measuring $m$ is also $\bra{e_m}\rho\ket{e_m}$.

If follows that instead of using a POVM of rank $1$, we can simply embed $\mathcal{H}_A$ in a space of dimension 
$d_A^2$ where $d_A = \dim \mathrm{supp}(\rho^A)$
 and make a full projective measurement.

We now assume $\mathcal{H}_A$ has been chosen of dimension $d_A^2$.
The optimum mutual information is then
\[
\sup_{U\in\mathrm{U}(\mathcal{H}_A)} I\big(\mathcal{D}\big((U\otimes \mathbf{1}_B)\rho(U^\dagger\otimes\mathbf{1})\big)^{A:B}\big)
\]
where $\mathrm{U}(\mathcal{H}_A) = \mathrm{U}(d_A^2)$ is the unitary group on $\mathcal{H}_A$.
Since $\mathrm{U}(d_A^2)$  is compact, since its action is continuous, and since 
$S(\rho^A)-S(\rho^{AB})$ is continuous in $\rho$ \cite{Alicki-Fannes} (and $U\otimes\mathbf{1}_B$ leaves $\rho^B$ fixed),
there is a $U$ for which the optimum is realized and $\sup$ may be replaced by $\max$.

\section{Zeroing conjugate off diagonal entries of  off diagonal blocks}\label{appb}
Since $I(\rho^{A:B})$ can never be less than $I(\mathcal{D}(\rho)^{A:B})$, one way to prove that equality
implies that $\rho^{AB}$ is block diagonal might be to show that if it were not, $I(\rho^{A:B})$ could be decreased
by zeroing non zero conjugate entries not on the block diagonal.  To proceed \cite{OZ} made the bold statement
that if conjugate non zero entries that are neither on the block diagonal, nor on any diagonal of the blocs, are 
replaced by $0$, then the entropy of the matrix strictly increases. That implies that $I(\rho^{A:B})$ decreases
since  then
$\rho^A$ and $\rho^B$ are left unchanged, only $S(\rho^{AB})$ is modified in
$I(\rho^{A:B}) = S(\rho^A) + S(\rho^B) - S(\rho^{AB})$.

Here is a Python $3$ program that takes a two qubit density operator
(thus a $4\times 4$ matrix comprising  four $2\times 2$ blocks), returns its eigenvalues and its Von Neumann
entropy and does the same on the matrix obtained after zeroing the $(00,11)$ and $(11,00)$ entries.
The  matrix has entropy $1.7555$ and after zeroing 
the two conjugate entries, the entropy decreases to $1.7546$ instead of increasing.
The entropy also decreases if we choose
the two other possible entries, $(01,10)$ and $(10,01)$. It is unclear how we could ever force the entropy to strictly
increase by such methods.
\footnotesize
\begin{verbatim}
from math import log, e
from numpy import array, linalg

def spec(m):
  return linalg.eigvalsh(m)
def S(m):
  sp = spec(m)
  return sum(-p*log(p,2) for p in sp if p !=0)
B = array([[ 0.25,   0.14,  -0.02,  -0.01],
           [ 0.14,   0.25,  -0.01,  -0.02],
           [-0.02,  -0.01,   0.25,   0.14],
           [-0.01,  -0.02,   0.14,   0.25]])
for i in range(2):
  print("The eigenvalues of\n %s" % (B))
  print("are  %s" % (spec(B)))
  print("with Von Neumann entropy %6.4f.\n" % (S(B)))
  B[0][3]=0
  B[3][0]=0
\end{verbatim}
\normalsize
\end{document}